\documentstyle[12pt,psfig]{article}
\begin{document}
\def \bb{\bibitem}
\def \nonumer{\nonumber}
\def \f{\frac}
\def \ra{\rightarrow}
\def \beq{\begin{equation}}
\def \eeq{\end{equation}}
\def \bea{\begin{eqnarray}}
\def \eea{\end{eqnarray}}
\def \ttl{\theta_l}
\def \thl{\theta_l}
\def \phil{\phi_l}
\def \ttt{\theta_t}
\def \tht{\theta_t}
\def \ee{e^+e^-}
\def \tbar{\overline{t}}
\def \ttbar{t\overline{t}}
\def \g{\gamma}
\def \re{{\rm Re}}
\def \im{{\rm Im}}
\vskip 1cm

\begin{center}
{\Large \boldmath \bf 
QCD corrections to decay-lepton polar and azimuthal angular distributions
in $e^+e^-\rightarrow t\overline{t}$
in the soft-gluon approximation 
}
\vskip 1cm
Saurabh D. Rindani
\vskip .2cm
{\it Theory Group, Physical Research Laboratory \\
 Navrangpura, Ahmedabad 380 009, India\\
\rm Email address: \tt saurabh@prl.ernet.in}
\end{center}
\vskip  1cm
\centerline{\small \bf Abstract}
\vskip .5cm

\begin{quote}
{\small
QCD corrections to order $\alpha_s$ in the soft-gluon approximation to angular
distributions of decay charged leptons in the process $\ee \ra \ttbar$, followed
by semileptonic decay of $t$ or $\tbar$, are obtained in the $\ee$
centre-of-mass frame. 
As compared to distributions in the top rest frame, these have the advantage 
that they would allow direct comparison
with experiment without the need to reconstruct the top rest frame. The results
also do not depend on the choice of a spin
quantization axis for $t$ or $\tbar$. Analytic expression for the 
triple distribution in the polar angle of $t$ and polar and 
 azimuthal angles  of the lepton is obtained. 
Analytic expression is also derived for the distribution in the charged-lepton
polar angle.  Numerical values are
discussed for $\sqrt{s}=400$ GeV, 800 GeV and 1500 GeV. 
}
\end{quote}

\vskip .5cm
\section{Introduction}

The discovery \cite{top} of a heavy top quark, with a mass of about 174 GeV
which is close to the electroweak symmetry breaking scale, raises the
interesting possibility that the study of its properties will provide hints to
the mechanism of symmetry breaking. 
While most of the gross properties
of the top quark will be investigated at the Tevatron and LHC, more accurate
determination of its couplings will have to await the construction of a linear
$e^+e^-$ collider. The prospects of the construction of such a linear collider,
which will provide detailed information also on the $W^{\pm}$, $Z$ and Higgs,
are under intense discussion currently, and it is very important at the present
time to focus on the details of the physics issues ( \cite{miller} and
references therein). 

In this context, top polarization is of great interest. There has been a lot of
work on 
production of polarized top quarks in the standard model (SM) 
in hadron \cite{pol1} collisions, and
$e^+e^-$ collisions in the continuum \cite{pol2}, as well as at the threshold
\cite{sum}.  A comparison of the theoretical
predictions for single-top polarization as well as $t\overline{t}$ spin
correlations with experiment can provide a verification of 
SM couplings and QCD corrections, or give clues to possible new physics
beyond SM in the couplings of the top quark [6-18]
(see \cite{sonireview}
for a review of CP violation in top physics).

Undoubtedly, the study of the top polarization is possible because of its large
mass, which ensures that the top decays fast enough for spin information not to
be lost due to hadronization \cite{kuhn}. Thus, kinematic distributions 
of top decay
products can be analysed to obtain polarization information. However, most
studies on top polarization, particularly on QCD corrections, have presumed an
accurate reconstruction of the spin, and are generally not concerned with decay
distributions. 
 Some works do consider decay distributions, but  
in the top rest frame \cite{jezabek,shadmi,kiyo}. 
In these cases, the onus of accurate determination of
the spin quantization axis and that of the top energy-momentum and hence of the
top rest frame, is largely left on the experimentalists.
The remaining papers which
discuss decay distributions in the laboratory frame are restricted to energy
distributions of decay products \cite{akatsu}, and do not calculate angular
distributions.

The choice of spin quantization axis has also been an issue of discussion
\cite{parke,shadmi}, and it has been remarked that a certain ``off-diagonal" 
spin basis has certain advantages.

	In an alternative approach, adopted usually in the context of the 
study of top polarization arising from new physics, predictions are made 
directly for decay-lepton \cite{bern,arens,pou0,pou,ter,chris2,grzad0,
grzad,sdr} 
or bottom-quark \cite{grzad,chris} distributions in the laboratory frame.
 Such an approach makes the issue of the choice of spin basis for the top quark 
superfluous. Moreover, if the study is restricted to energy and polar angle 
distributions of top decay products, it even obviates the need for accurate 
determination of the energy or momentum direction of the top quark \cite{pou}.
Even in case azimuthal distributions are studied, where an additional direction
is needed to define the azimuthal angle, the knowledge of the
magnitude of the top momentum is not required. It is sufficient if the direction
is determined with reasonable accuracy.

In this paper we shall be concerned with the laboratory-frame angular 
distribution of secondary leptons arising from the decay of the top quarks in 
$e^+e^- \rightarrow t\overline{t}$ in the context of QCD corrections to 
order $\alpha_s$. QCD corrections to top polarization in $e^+e^- \ra \ttbar$
have been  calculated earlier by many groups \cite{kiyo}-\cite{liu}.
In cases when the corrections have been applied to the process 
including top 
decay, the discussion of angular distributions is  restricted to the
top rest frame \cite{kiyo}. QCD corrections to the lepton energy
distributions have been treated in the top rest frame in 
\cite{jezabek} and in the lab. frame in \cite{akatsu}. This 
paper provides, for the first time, angular distribution in the $e^+e^-$ 
centre-of-mass (c.m.) frame. As a first approach, this work is restricted, for 
simplicity, to the soft-gluon approximation (SGA). SGA has been found to give a 
satisfactory description of top polarization in single-top production 
\cite{kodaira}, and it is hoped that it will suffice to give a reasonable
quantitative description.

The study of the laboratory (lab.) frame angular distribution of secondary 
leptons, besides admitting direct experimental observation,
 has another advantage. It has been found \cite{grzad, sdr} that the 
angular distribution is not altered, to first-order approximation, 
by modifications of the $tbW$ decay vertex, provided the $b$-quark mass 
is neglected. Thus, our result would hold to a high degree of accuracy even 
when ${\cal O}(\alpha_s )$ soft-gluon QCD corrections to top decay are included,
since these can be represented by the same form factors \cite{lampe2} 
considered in \cite{grzad,sdr}. We do 
not, therefore, need to calculate these explicitly. It is sufficient to include 
${\cal O}(\alpha_s )$ corrections to the $\gamma t\overline{t}$ and 
$Z t\overline{t}$ vertices. This, of course, assumes that QCD corrections of 
the nonfactorizable type \cite{been}, where a virtual gluon is exchanged 
gluon between 
$t$ ($\overline{t}$) and  $\overline{b}$ ($b$) from $\overline{t}$ ($t$) decay,
 can be neglected. We have assumed that these are negligible.

The procedure adopted here is as follows. We make use of effective $\gamma 
t\overline{t}$ and  $Z t\overline{t}$ vertices derived in earlier works in the 
soft-gluon approximation, using an appropriate cut-off on the soft-gluon energy.
 In principle, these effective vertices are obtained by suitably cancelling the 
infra-red divergences in the virtual-gluon contribution to the differential 
cross section for $e^+e^- \rightarrow t\overline{t}$ against the real soft-gluon
 contribution to the differential cross section for $e^+e^- \rightarrow 
t\overline{t} g$. For practical purposes, restricting to SGA, it is sufficient 
to modify the tree-level $\gamma t\overline{t}$ and $Z t\overline{t}$ vertices 
suitably to produce the desired result. Thus, assuming ${\cal O}(\alpha_s )$ 
effective SGA vertices, we have obtained  helicity amplitudes for $e^+e^- 
\rightarrow t\overline{t}$, and hence spin-density matrices for production. 
This implies an assumption that these effective vertices provide, in SGA, a 
correct approximate description of the off-diagonal density matrix elements as 
well as the diagonal ones entering the differential cross sections. 
Justification for this would need explicit calculation of hard-gluon effects, 
and is beyond the scope of this work.

We have considered three possibilities, corresponding to the electron beam
being unpolarized ($P=0$), fully left-handed polarized ($P=-1$), and fully
right-handed polarized ($P=+1$). Since we give explicit analytical expressions,
suitable modification to more realistic polarizations would be straightforward.

The results for the case of polar distributions of the decay lepton, restricted
to $\sqrt{s}$ values of 400 and 800 GeV, were reported earlier in a short paper
\cite{PLB}. We present here more details, as well as azimuthal distributions.
We also consider the possibility of a higher energy option of the linear
collider, with $\sqrt{s}=1500$ GeV.

Our main result may be summarized as follows. By and large the distribution in
the polar angle $\theta_l$ of the secondary lepton w.r.t. the $e^-$ beam
direction is unchanged in shape on inclusion of QCD corrections in SGA. The
$\theta_l$ distribution for $\sqrt{s}=400$ GeV 
is very accurately described by overall
multiplication by a $K$ factor ($K\equiv 1+\kappa >1$), 
except for extreme values of
$\theta_l$, and that too for the case of $P=+1$. For the other energies
considered, $\kappa$ continues to be slowly varying function of $\thl$. 
This has the
important consequence that earlier results  on the sensitivity of lepton
angular distributions or asymmetries to anomalous top couplings, obtained for
$\sqrt{s}$ values around 400 GeV
without QCD corrections being taken into account, would go through by a simple
modification by a factor of $1/\sqrt{K}$ \cite{kek}.
The triple distributions in $\theta_t$, $\theta_l$ and $\phi_l$  show an
asymmetry around $\phil = 180^{\circ}$, which is not present at Born level.

\section{Derivation of angular distributions to ${\cal O}(\alpha_s)$}

We first obtain expressions for helicity amplitudes for 
\beq 
e^-(p_{e^-}) + e^+(p_{e^+}) \rightarrow t(p_t) + \overline{t}(p_{\overline{t}})
\eeq
going through virtual $\gamma$ and $Z$ in the $\ee$ c.m. frame, including QCD
corrections in SGA. The starting point is the QCD-modified $\gamma \ttbar$ and
$Z\ttbar$ vertices obtained earlier (see, for example, \cite{kodaira, tung}). 
We can write them 
\cite{kodaira}
in the limit of vanishing electron mass as 
\beq\label{gvert}
\Gamma_{\mu}^{\gamma} = e \left[ c_v^{\gamma} \gamma_{\mu} + c_M^{\gamma}
\frac{(p_t - p_{\overline{t}})_{\mu}}{2 m_t} \right],
\eeq
\beq\label{Zvert}
\Gamma_{\mu}^{Z} = e \left[ c_v^{Z} \gamma_{\mu} +  c_a^{Z} \gamma_{\mu}\gamma_5  + c_M^{Z}
\frac{(p_t - p_{\overline{t}})_{\mu}}{2 m_t} \right],
\eeq
where
\beq
c_v^{\gamma} = \f{2}{3} (1+A),
\eeq
\beq
c_v^Z = \frac{1}{\sin\theta_W \cos\theta_W} \left( \frac{1}{4} - \frac{2}{3}
\sin^2\theta_W\right) (1+A),
\eeq
\beq
c_a^{\gamma}=0,
\eeq
\beq
c_a^Z = \frac{1}{\sin\theta_W \cos\theta_W} \left(-\frac{1}{4}\right) (1+A+2 B),
\eeq
\beq
c_M^{\gamma} = \frac{2}{3} B,
\eeq
\beq
c_M^Z = \frac{1}{\sin\theta_W \cos\theta_W} \left( \frac{1}{4} - \frac{2}{3}
\sin^2\theta_W\right) B.
\eeq
The form factors $A$ and $B$ are given to order $\alpha_s$ in SGA by 
\bea\label{A}
{\rm Re} A &=& \hat{\alpha}_s \left[ \left( \frac{1+\beta^2}{\beta} \log
\frac{1+\beta}{1-\beta} - 2\right)\log\frac{4 \omega^2_{\rm max}}{m_t^2} - 4
\right.\nonumber \\
&& \left. + \frac{2+3\beta^2}{\beta}\log\frac{1+\beta}{1-\beta} +
\frac{1+\beta^2}{\beta} \left\{ \log\frac{1-\beta}{1+\beta}\left( 3
\log\frac{2\beta}{1+\beta}\right.\right.\right. \nonumber \\
&& \left.\left.\left. + \log\frac{2\beta}{1-\beta} \right) + 4 {\rm Li}_2
\left(\frac{1-\beta}{1+\beta}\right) + \frac{1}{3}\pi^2\right\}\right],
\eea
\beq\label{B}
{\rm Re} B=\hat{\alpha}_s \frac{1-\beta^2}{\beta} \log \frac{1+\beta}{1-\beta},
\eeq
\beq
{\rm Im} B = - \hat{\alpha}_s \pi \frac{1-\beta^2}{\beta},
\eeq
where $\hat{\alpha}_s=\alpha_s/(3\pi)$, $\beta = \sqrt{1-4 m_t^2/s}$, and Li$_2$
is the Spence function. ${\rm Re} A$ in eq. (\ref{A}) contains the effective 
form factor
for a cut-off $\omega_{\rm max}$ on the gluon energy after the infrared
singularities have been cancelled between the virtual- and soft-gluon
contributions in the on-shell renormalization scheme. Only the real part of
the form factor $A$ has been given, because the contribution of the imaginary 
part is proportional to the $Z$ width, and hence negligibly small
\cite{ravi,kodaira}. The imaginary part of $B$, however, contributes to the
triple angular distribution.

The vertices in eqs. (\ref{gvert}) and (\ref{Zvert}) can be used to obtain
helicity amplitudes for $\ee \rightarrow \ttbar $, including the contribution of
$s$-channel $\gamma$ and $Z$ exchanges. The result is, in a notation where the
subscripts of $M$ denote the signs of the helicities of $e^-$, $e^+$, $t$ and
$\overline{t}$, in that order, 
\beq
M_{+-\pm \pm} = \pm \frac{4e^2}{s}\sin\ttt \f{1}{\gamma} \left[
\left(c_v^{\gamma} + r_R c_v^Z \right) - \beta^2 \gamma^2 \left( c_M^{\gamma} +
r_R c_M^{Z} \right) \right],
\eeq
\beq
M_{-+\pm \pm} = \pm \frac{4e^2}{s}\sin\ttt \f{1}{\gamma} \left[
\left(c_v^{\gamma} + r_L c_v^Z \right) - \beta^2 \gamma^2 \left( c_M^{\gamma} +
r_L c_M^{Z} \right) \right],
\eeq
\beq
M_{+-\pm \mp} = \frac{4e^2}{s} \left( 1 \pm \cos\ttt \right) \left[\pm
\left(c_v^{\gamma} + r_R c_v^Z \right) +  \beta \left(c_a^{\gamma} + r_R c_a^Z
\right)\right],
\eeq
\beq
M_{-+\pm \mp} = \frac{4e^2}{s} \left( 1 \mp \cos\ttt \right) \left[\mp
\left(c_v^{\gamma} + r_L c_v^Z \right) -  \beta \left(c_a^{\gamma} + r_L c_a^Z
\right)\right],
\eeq
where $\ttt$ is the angle the top-quark momentum makes with the $e^-$ momentum,
$\gamma= 1/\sqrt{1-\beta^2}$, and $r_{L,R}$ are related to the left- and
right-handed $Ze\overline{e}$ couplings, and are given by
\beq
r_L = \left(\f{s}{s-m_Z^2}\right) \f{1}{\sin\theta_W\cos\theta_W},
\eeq
\beq
r_R = -\left(\f{s}{s-m_Z^2}\right) \tan\theta_W.
\eeq

Since we are interested in lepton distributions arising from top decay, we also
evaluate the helicity amplitudes for $t\rightarrow b l^+ \nu_l$ ( or $
\overline{t} \rightarrow \overline{b} l^- \overline{\nu}_l$), which will be
combined with the production amplitudes in the narrow-width approximation for
$t$ ($\overline{t}$) and $W^+$ ($W^-$). In principle, QCD corrections should be
included also in the decay process. However, in SGA, these could be written in
terms of effective form factors \cite{lampe2}. 
As found earlier \cite{grzad, sdr}, in the
linear approximation, these form factors do not affect the charged-lepton
angular distribution. Hence we need not calculate these form factors.
Nevertheless, to illustrate this point we will discuss an arbitrary
$\overline{t}bW$ vertex, in the limit of vanishing bottom mass.

The most general $\overline{t}bW$ vertex may be written as
\bea
\Gamma^{\mu}_{t\rightarrow bW}& =& - \f{g}{\sqrt{2}} V_{tb} \overline{u}(p_b)
\left[ \gamma^{\mu} (f_{1L} P_L + f_{1R} P_R)\right. \nonumber \\
&& \left. -\f{i}{m_W}\sigma^{\mu\nu}
(p_t-p_b)_{\nu} (f_{2L}P_L + f_{2R} P_R)\right] u(p_t),
\eea
where $P_{L,R} = \f{1}{2}(1\mp\gamma_5)$, and $V_{tb}$ is the
Cabibbo-Kobayashi-Maskawa (CKM) matrix element, which we will take to be equal
to one. $f_{1R}$ and $f_{2L}$ do not contribute in the limit of vanishing $b$
mass, and can be omitted. Moreover, we will set $f_{1L}=1$, the value at tree
level. The result does not depend on the value of $f_{1L}$, as we will see
soon. The decay helicity amplitudes in the $t$ rest frame can be found in
\cite{sdr}, and we do not repeat them here. 

It was found in \cite{sdr} that, in
this case, the spin density matrix for $t\rightarrow bl^+\nu_l$, on integration
over the $b$-quark azimuthal angle, is the same as
that for the Born vertex, except for an overall factor $\left(1+{\rm Re}f_{2R}
\frac{m_Wm_t}{p_t\cdot p_{l^+}}\right)$ multiplying all the matrix elements.
Moreover, after performing a boost to the lab. frame, and
integrating over the charged-lepton energy, that the $f_{2R}$ dependent factor
cancelled against the corresponding factor in the correction to the top width.
If we had not set $f_{1L}=1$, the $f_{1L}$ dependence would also have likewise
cancelled. 

The final result for the angular distribution in the lab. frame can be written
as  
\bea\label{triple}
\lefteqn{\f{d^3\sigma}{d\cos\ttt d\cos\thl d\phil} = \f{3\alpha^2\beta
m_t^2}{8s^{2}}B_l \f{1}{(1-\beta\cos\theta_{tl})^3}} 
\nonumber\\
&\times &\left[{\cal  A} (1-\beta\cos\theta_{tl})
+{\cal B} (cos\theta_{tl} - \beta )\right. \nonumber \\
&&\left.
+{\cal  C} (1-\beta^2)\sin\tht\sin\thl (\cos\tht\cos\phil - 
\sin\tht\cot\thl )\right. \nonumber \\
&&\left. +{\cal D} (1-\beta^2) \sin\tht\sin\thl\sin\phil \right],
\eea
where $\tht$ and $\thl$ are polar angles of respectively of the $t$ and $l^+$
momenta with respect to the $e^-$ beam direction chosen as the $z$ axis, and
$\phil$ is the azimuthal angle of the $l^+$ momentum relative to an axis chosen
in the $\ttbar$ production plane. $B_l$ is the leptonic branching ratio of the
top. $\theta_{tl}$ is the angle between the $t$ and
$l^+$ directions, given by
\beq
\cos\theta_{tl} = \cos\tht\cos\thl + \sin\tht\sin\thl\cos\phil ,
\eeq
and the coefficients ${\cal A}$, ${\cal B}$, ${\cal C}$ and ${\cal D}$ are
given  by
\bea
{\cal A} & = & A_0 + A_1 \cos\tht + A_2 \cos^2\tht , \\
{\cal B} & = & B_0 + B_1 \cos\tht + B_2 \cos^2\tht , \\
{\cal C} & = & C_0 + C_1 \cos\tht , \\
{\cal D} & = & D_0 + D_1 \cos\tht ,
\eea
with \cite{fn}
\bea
A_0 & = & 2 \left\{  (2-\beta^2) \left[ 2\vert c_v^{\g}\vert ^2 +
2(r_L+r_R){\rm Re}\left(c_v^{\g}c_v^{Z*}\right)
 + (r_L^2+r_R^2)\vert c_v^Z\vert ^2 \right] 
\right.  \nonumber \\
&& \left. 
+  \beta^2 (r_L^2+r_R^2)\vert c_a^Z \vert ^2 
-2\beta^2 \re \left[2c_v^{\g}c_M^{\g *}
+(r_L+r_R)\left( c_v^{\g}c_M^{Z*}+c_v^Zc_M^{\g *} \right)
\right. \right. \nonumber \\
&& \left. \left. + (r_L^2+r_R^2) c_v^Zc_M^{Z*} \right]
\right\} (1- P_e P_{\overline e}) 
+ 2 \left\{ 
    \beta^2 (r_L^2-r_R^2)\vert c_a^Z\vert ^2
\right. \nonumber \\  & & \left.
+(2-\beta^2) \left[
2(r_L-r_R)\re \left(c_v^{\g}c_v^{Z*}\right) + (r_L^2-r_R^2)\vert c_v^Z \vert
^2 \right] 
\right.  \nonumber \\
&& \left.
- 2\beta^2 \re
\left[(r_L-r_R)\left(c_v^{\g}c_M^{Z*}+c_v^Zc_M^{\g *}\right) 
 + (r_L^2 - r_R^2)  c_v^Zc_M^{Z*}  
\right] \right\} (P_{\overline e}-P_e), \nonumber \\ 
A_1&=& -8 \beta \re \left( c_a^{Z*} \left\{
\left[(r_L-r_R)c_v^{\g} + (r_L^2-r_R^2)c_v^Z \right]  (1-
P_eP_{\overline e}) \right.\right. \nonumber \\ && \left. \left. +
\left[(r_L+r_R)c_v^{\g} + (r_L^2+r_R^2)c_v^Z \right] (P_{\overline
e}-P_e) \right\} \right),  \nonumber\\ 
A_2&=& 2 \beta^2 \left\{ \left[
2\vert c_v^{\g}\vert ^2 + 4\re \left( c_v^{\g}c_M^{\g *}\right) 
+ 2(r_L+r_R)\re \left(c_v^{\g}c_v^{Z*} +c_v^{\g}c_M^{Z*} 
+ c_v^Zc_M^{\g *}\right)  
\right.\right. \nonumber \\
&&\left. \left. 
+(r_L^2+r_R^2)\left(
\vert c_v^Z \vert ^2 + \vert c_a^Z \vert ^2 +2 \re \left(c_v^Zc_M^{Z*} \right)
\right)  \right]  (1- P_e P_{\overline e})
\right. \nonumber \\ && \left. +  \left[ 2(r_L-r_R)\re \left(c_v^{\g}c_v^{Z*}
+c_v^{\g}c_M^{Z*} 
+ c_v^Z c_M^{\g *}\right) \right. \right. \nonumber \\
&& \left.\left.
+(r_L^2-r_R^2)\left( \vert c_v^Z \vert ^2 +  \vert c_a^Z \vert ^2 
+2 \re \left( c_v^Zc_M^{Z*}\right) \right) \right]
(P_{\overline e}-P_e) \right\} , \nonumber\\ 
B_0&=& 4\beta
\left\{ r_L \re  \left[ \left(c_v^{\g} + r_Lc_v^Z\right)  c_a^{Z*} \right] 
 (1-P_e)(1+P_{\overline e})
\right. \nonumber \\ && \left.+ r_R \re \left[ \left(c_v^{\g} + r_Rc_v^Z\right)
c_a^{Z*} \right]
(1+P_e)(1-P_{\overline e}) \right\} , \nonumber \\ 
B_1&=& -4 \left\{
\left[\vert c_v^{\g}+r_Lc_v^Z\vert ^2+ \beta^2 r_L^2 \vert c_a^Z \vert ^2 
\right]
(1-P_e)(1+P_{\overline e}) \right. \nonumber \\ && \left.-
\left[\vert c_v^{\g}+r_Rc_v^Z\vert ^2+ \beta^2 r_R^2\vert c_a^Z\vert ^2 \right]
(1+P_e)(1-P_{\overline e}) \right\} , \nonumber \\ 
B_2&=& 4\beta
\left\{r_L \re \left[ \left(c_v^{\g} + r_Lc_v^Z\right)  c_a^{Z*} \right] 
 (1-P_e)(1+P_{\overline e})
\right. \nonumber \\ && \left.
+ r_R \re \left[ \left(c_v^{\g} + r_Rc_v^Z\right)
c_a^{Z*} \right]
(1+P_e)(1-P_{\overline e}) \right\} , \nonumber\\ 
C_0&=& \!\!\! 4
\left\{ \vert c_v^{\g}\! +\! r_Lc_v^Z \vert ^2\! - \!\beta^2 \g^2
\re\! \left[ \left( c_v^{\g} + r_L c_v^Z\right)\!\! \left( c_M^{\g *}+r_L c_M^{Z*}
 \right)\right] \right\} (1-P_e)(1+P_{\overline e}) \nonumber \\
& & \!\!\!\!\! - \! 4\left\{ \vert c_v^{\g}\! +\! r_R c_v^Z\vert ^2\! - \!
\beta^2
\gamma^2 \re \!\left[ \left( c_v^{\g} + r_R c_v^Z\right)\!\! \left(  c_M^{\g *}
 + r_R  c_M^{Z *}
\right)\right] \right\} (1+P_e)(1-P_{\overline e}) , \nonumber\\
C_1&=&- 4\beta \re \left\{ \left[ \left( c_v^{\g} +
r_Lc_v^Z\right)   - \beta^2\gamma^2\left( c_M^{\g} + r_L
c_M^Z \right)\right]r_L c_a^{Z*}  (1-P_e)(1+P_{\overline e}) \right. 
\nonumber \\ && \left.  + \left[ \left( c_v^{\g} + r_Rc_v^Z\right)   
- \beta^2\gamma^2 \left(c_M^{\g} + r_R c_M^Z \right)
\right] r_R c_a^{Z*} (1+P_e)(1-P_{\overline e})
\right\},\nonumber \\
D_0&=&4 \beta  \im \left[ \left\{ \left (c_v^{\g} + r_L c_v^Z 
\right) -
\beta^2 \g^2 \left( c_M^{\g} + r_L c_M^Z \right) \right\}
r_L c_a^{Z*} (1-P_e)(1+P_{\overline e}) \right. \nonumber \\
&&- \left. 
\left\{ \left(c_v^{\g} + r_R c_v^Z \right) -
\beta^2 \g^2 \left( c_M^{\g} + r_R c_M^Z \right) \right\}
r_R c_a^{Z*}(1+P_e)(1-P_{\overline e}) \right] \sin\theta_t,
\nonumber \\
D_1&=&4\beta^2\g^2  \im \left[ \left(c_M^{\g}+r_L c_M^Z \right)
\left( c_v^{\g *} + r_L c_v^{Z*} \right) (1-P_e)(1+P_{\overline e})\right.
\nonumber \\
&& + \left. \left(c_M^{\g}+r_R c_M^Z \right)
\left( c_v^{\g *} + r_R c_v^{Z*} \right) (1+P_e)(1-P_{\overline e})\right] 
\sin\theta_t.
 \nonumber 
\eea

Integrating over $\phil$ and $\theta_{tl}$ we get the $\thl$ distribution:
\bea
\f{d\sigma}{d\cos\thl} & = & \f{3\pi\alpha^2}{32s}\beta B_l\left\{
\left(4A_0 +
\f{4}{3}A_2\right) + \left[ -2A_1
\left(\f{1-\beta^2}{\beta^2}\log\f{1+\beta}{1-\beta}
-\f{2}{\beta}\right)\right. \right.\nonumber \\
&&\left.\left. + 2 B_1 \f{1-\beta^2}{\beta^2} \left( \f{1}{\beta} 
\log\f{1+\beta}{1-\beta} - 2 \right) \right.\right. \nonumber \\
&& \left. \left. + 2C_0 \f{1-\beta^2}{\beta^2} 
\left( \f{1-\beta^2}{\beta} \log\f{1+\beta}{1-\beta} - 2 \right) \right]
\cos\thl \right. \nonumber \\
&& \left. + \left[ 2A_2 \left( \f{1-\beta^2}{\beta^3} \log\f{1+\beta}{1-\beta}
-\f{2}{3\beta^2}\left(3-2\beta^2\right)\right) \right. \right.
\nonumber\\
&& \left. \left. + \f{1-\beta^2}{\beta^3} \left\{ B_2 
\left( \f{\beta^2-3}{\beta} \log
\f{1+\beta}{1-\beta} + 6 \right) \right. \right. \right. \nonumber \\
&& \left. \left.\left. - C_1 \left( \f{3(1-\beta^2)}{\beta}
\log \f{1+\beta}{1-\beta} - 2 (3-2 \beta^2)\right)\right\} \right] \right.
\nonumber \\
&& \left. \times 
 (1-3\cos^2 \thl)\right\}.
\eea

\begin{figure}[ptb]
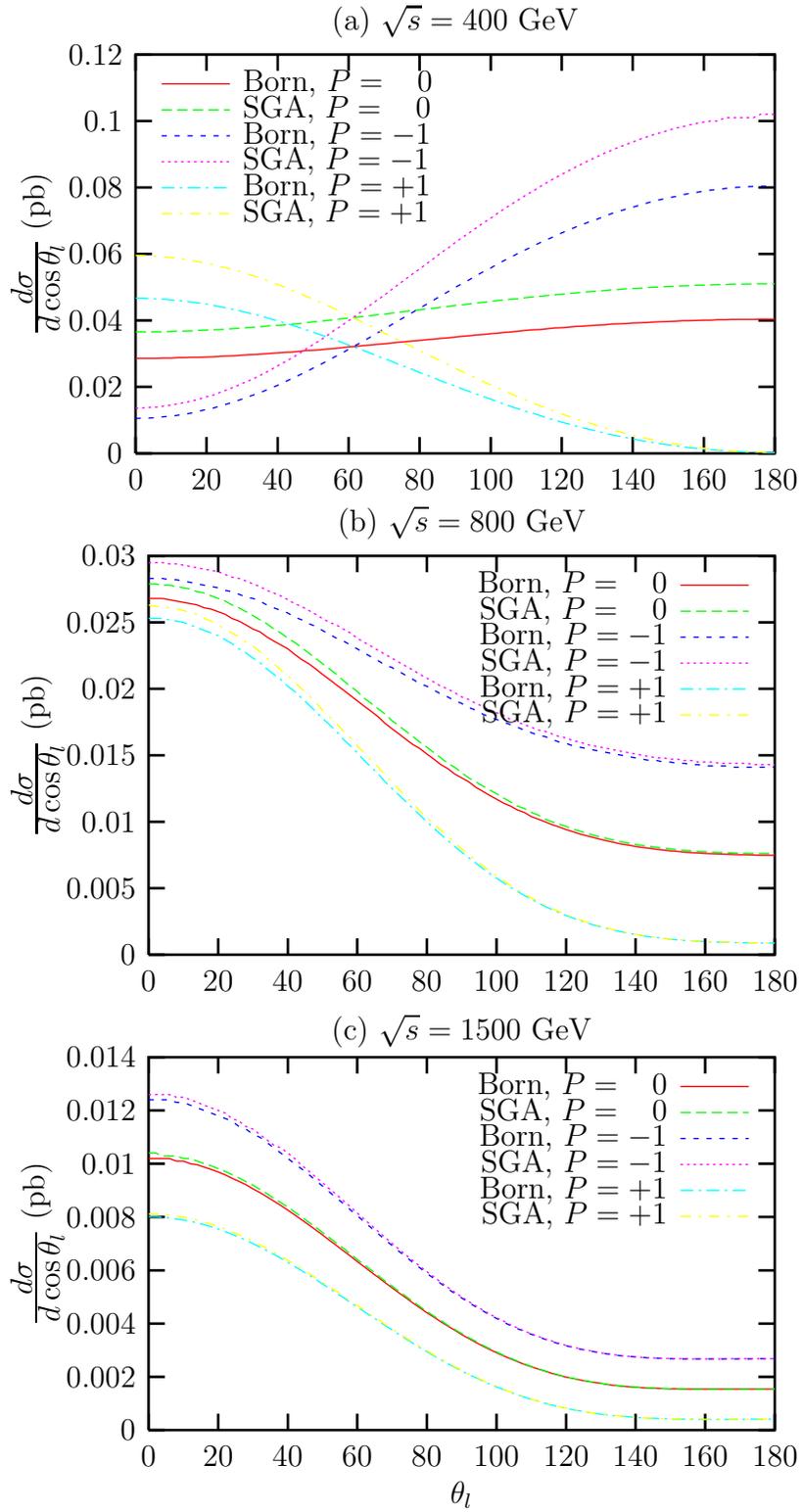

\begin{center}
\vskip -3.81cm
\input{angdist1.tex}
\vskip .3cm
\input{angdist2.tex}
\vskip .3cm
\input{angdist3.tex}
\vskip -.3cm
\caption{\small 
The distribution in $\theta_l$ with and without QCD corrections for
(a) $\sqrt{s}=400$ GeV, (b) $\sqrt{s}=800$ GeV and (c) $\sqrt{s}=1500$ GeV
plotted against $\theta_l$, for $e^-$ beam polarizations $P=0,-1,+1$ in each
case.}
\end{center}
\label{graph1}
\end{figure}

%\centerline{\protect\hbox{\psfig{file=graph1.ps,height=7cm}}}

\section{Numerical Results}

After having obtained analytic expressions for angular distributions, we now
examine the numerical values of the QCD corrections.

We use the parameters $\alpha = 1/128$, $\alpha_s(m_Z^2)=0.118$, $m_Z= 91.187$
GeV, $m_W=80.41$ GeV, $m_t=175$ GeV and $\sin^2\theta_W=0.2315$.  We consider
leptonic decays into one specific channel (electrons or muons or tau leptons), 
corresponding to a branching ratio of $1/9$. We have used,
following \cite{kodaira}, a gluon energy cut-off of 
$\omega_{\rm max}=(\sqrt{s}-2m_t)/5$. While qualitative results would be
insensitive, exact
quantitative results would of course depend on the choice of cut-off.

In Figs. 1 (a)-(c) we show the single differential cross section
$\f{d\sigma}{d\cos\thl}$ in picobarns with and without QCD corrections, 
for three values of
$\sqrt{s}$, viz., (a) 400 GeV, (b) 800 GeV and (c) 1500 GeV, and for $e^-$ beam
polarizations $P=0,-1,+1$. It can be seen
that the distribution with QCD corrections follows, in general,  the shape of
the lowest order distribution.

\begin{figure}[ptb]
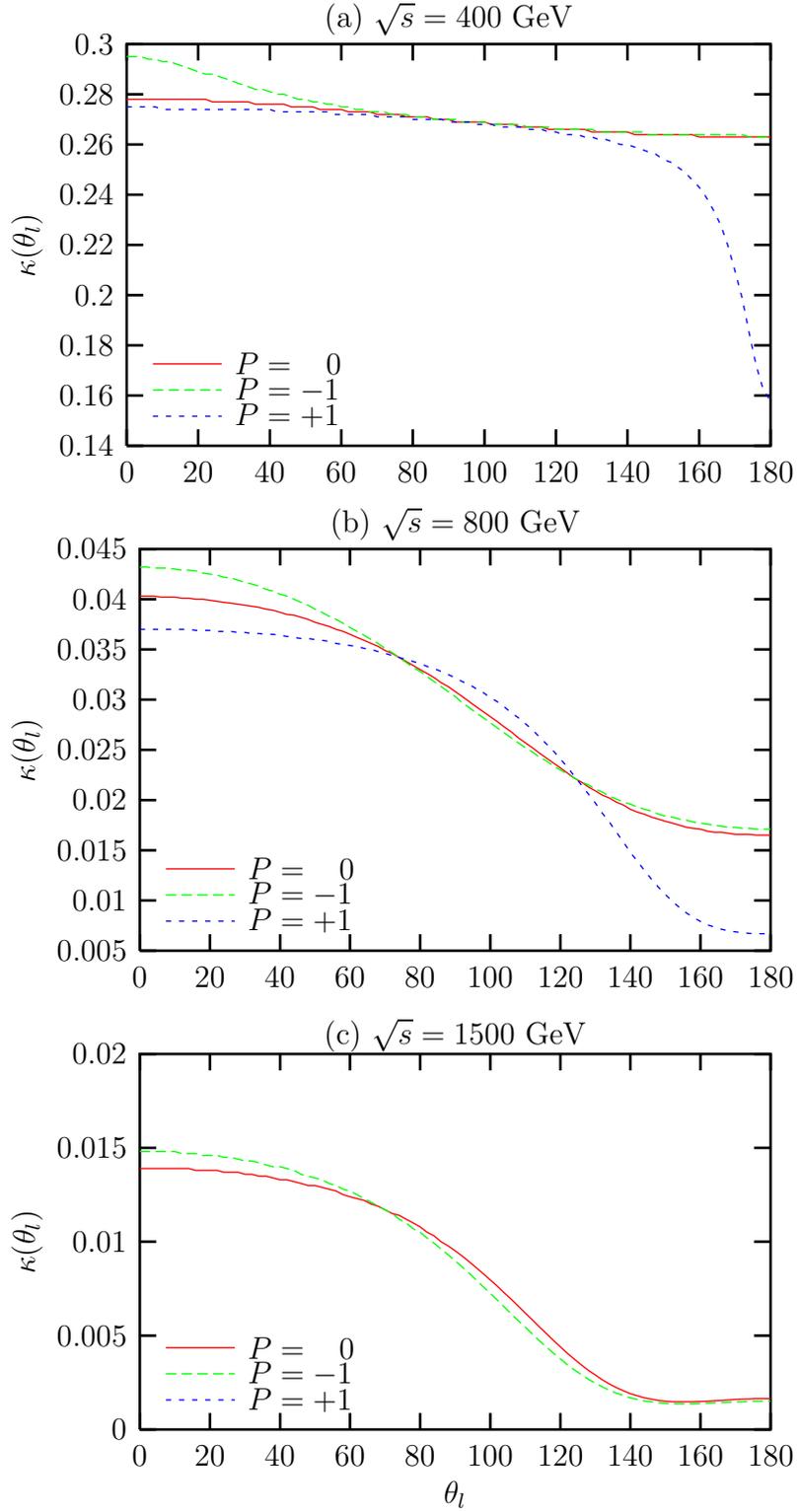

\begin{center}
\vskip -3.81cm
\input{delangdist1.tex}
\vskip .3cm
\input{delangdist2.tex}
\vskip .3cm
\input{delangdist3.tex}
\vskip -.3cm
\caption{\small 
The fractional QCD contribution $\kappa (\theta_l)$ defined in the
text for
(a) $\sqrt{s}=400$ GeV, (b) $\sqrt{s}=800$ GeV and (c) $\sqrt{s}=1500$ GeV
plotted as a function of $\theta_l$, for $P=0,-1,+1$.}
\label{graph2}
\end{center}
\end{figure}

In Figs. 2 (a)-(c) is displayed the fractional deviation of the QCD-corrected
distribution from the lowest order distribution:
\beq
\kappa (\thl) = \left( \f{d\sigma_{Born}}{d\cos\thl}\right) ^{-1} 
\left( \f{d\sigma_{SGA}}{d\cos\thl}
-\f{d\sigma_{Born}}{d\cos\thl} \right) .
\eeq
It can be seen that $\kappa(\thl)$ is independent of $\thl$ to a fair degree of
accuracy for $\sqrt{s}=400$ GeV.

In Figs. 3 (a)-(c) we show the fractional QCD contributions 
$(F_{\rm SGA} - F_{\rm Born})/F_{\rm Born}$
where $F(\theta_l)$,
is the normalized distribution:
\beq
F(\theta_l)  = \f{1}{\sigma}\left( \f{d\sigma}{d\cos\thl} \right) .
\eeq
It can be seen that the fractional change in the normalized distributions for
$\sqrt{s}=400$ GeV is at most of the order of 1 or 2\% (except in the case of
$P=+1$, for $\thl \geq 160^{\circ}$). For the other values of $\sqrt{s}$, it is
even smaller. This implies that QCD corrected angular
distribution is well approximated, at the per cent level, by a constant
rescaling by a $K$ factor.

\begin{figure}[ptb]
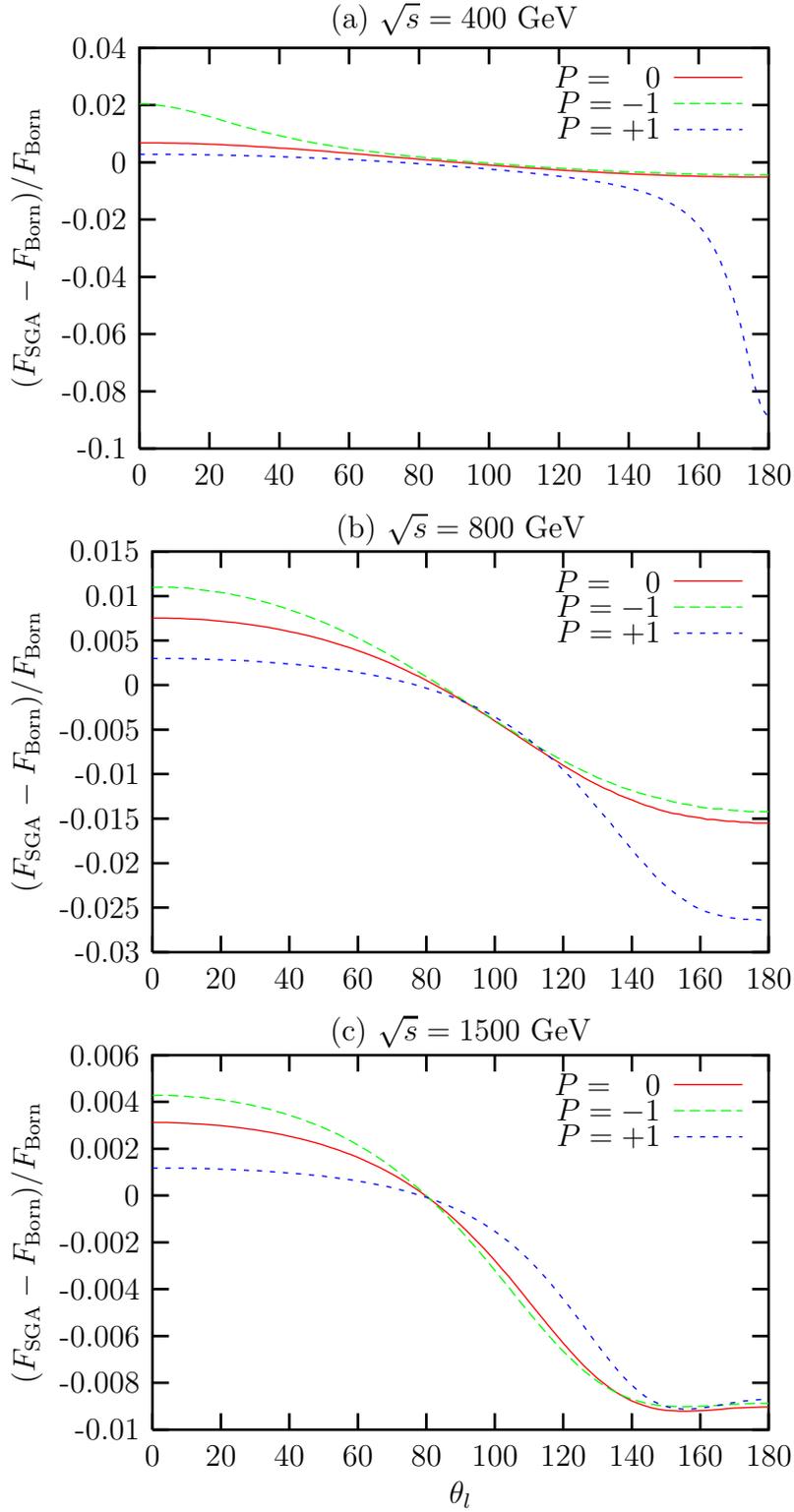

\begin{center}
\vskip -3.81cm
\input{normangdist1.tex}
\vskip .3cm
\input{normangdist2.tex}
\vskip .3cm
\input{normangdist3.tex}
\vskip -.32cm
\caption{\small 
The fractional QCD contribution in normalized angular distributions, 
$F(\theta_l)$ defined in the
text,  for
(a) $\sqrt{s}=400$ GeV, (b) $\sqrt{s}=800$ GeV and (c) $\sqrt{s}=1500$ GeV
plotted as a function of $\theta_l$, for $P=0,-1,+1$.}
\end{center}
\end{figure}

We now discuss the triple differential cross section, which corresponds to a
distribution in $\tht$, $\thl$ and $\phil$, for which we have obtained an
analytic expression, eq. (\ref{triple}). This distribution, unlike the $\thl$
distribution, requires the knowledge of the direction of $t$. In fact,
without knowing the direction of the top momentum, it is not even possible to
define $\phil$. However, the magnitude of the momentum need not be determined 
precisely.

We have plotted in Fig. 4(a) $d^3\sigma/(d\cos\tht d\cos\thl d\phil )$,
including QCD correction, against $\phil$ for different values of $\thl$, for a
fixed value of $\tht = 30^{\circ}$, for $\sqrt{s}=400$ GeV, and for $P=0$. Fig.
4(b) gives the corresponding fractional deviations from the lowest-order values,\beq
\kappa(\theta_t,\thl,\phil)=\left[  
\f{d^3\sigma_{\rm Born}}{d\cos\tht d\cos\thl
d\phil}\right]^{-1}\left[ \f{d^3\sigma_{\rm SGA}}{d\cos\tht d\cos\thl 
d\phil} - \f{d^3\sigma_{\rm Born}}{d\cos\tht d\cos\thl d\phil}\right]\!.
\eeq
It can be seen that the QCD contribution remains around 28\% for various values
of $\thl$ and $\phil$. 
The triple distributions show an
asymmetry around $\phil = 180^{\circ}$, which is not present at Born level.
This asymmetry is not visible in the Fig. 4 (a), but is very clear in Fig.
4 (b), where fractional QCD contributions to the normalized distributions are 
plotted.

The same plots as in Fig. 4 (a), (b) are repeated for $\sqrt{s}=800$ GeV in
Figs. 5 (a), (b) and for $\sqrt{s}=1500$ GeV in Figs. 6 (a), (b).

\begin{figure}[ptb]
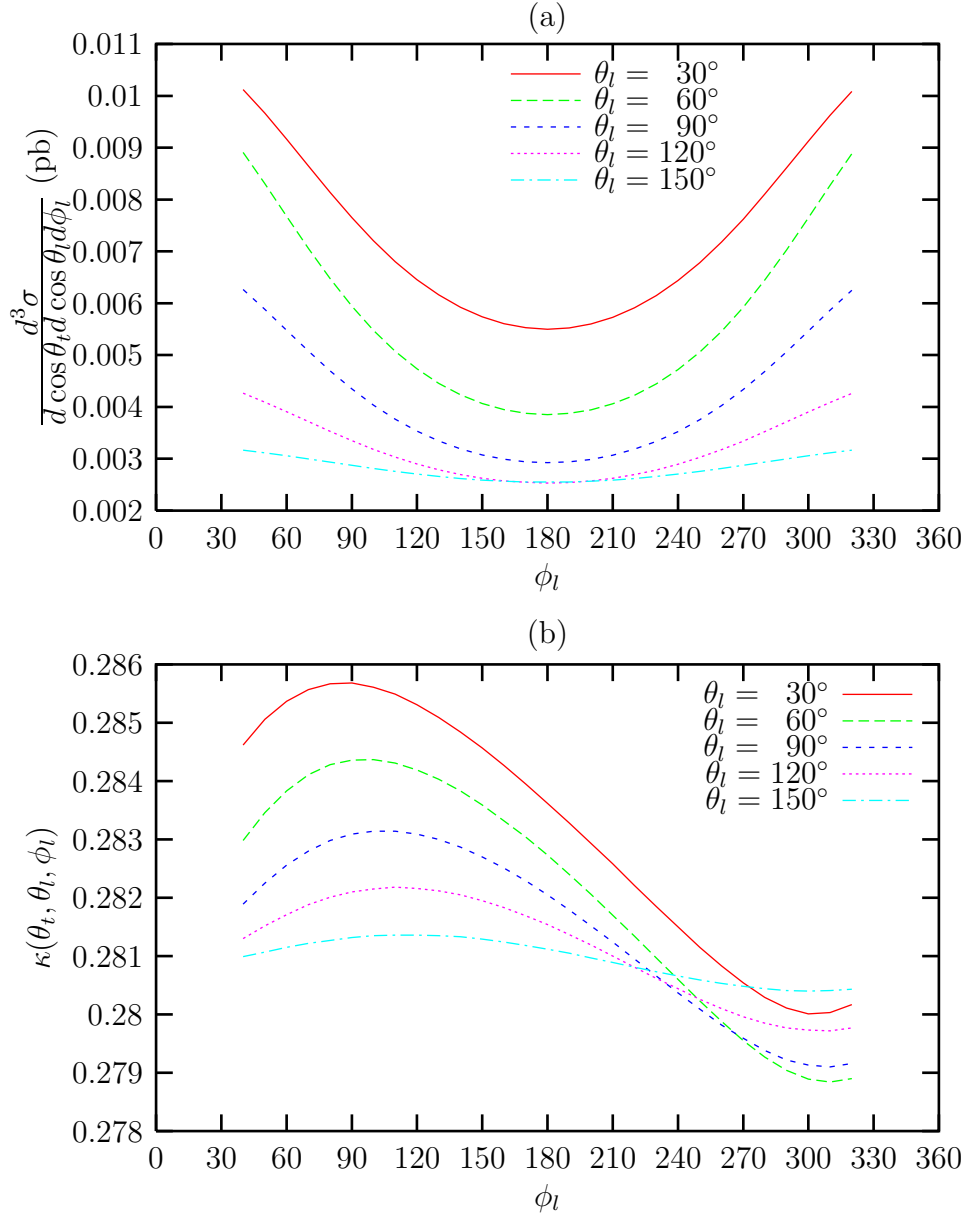

\begin{center}
\vskip -1.7cm
\input{dist3011.tex}
\vskip .6cm
\input{dist3012.tex}
\caption{ (a) The triple differential distribution including the  
QCD contribution 
for $\sqrt{s}=400$ GeV and $\tht=30^{\circ}$. 
(b) The fractional QCD contribution $\kappa(\tht,\thl,\phil)$ to the triple 
differential distribution 
for $\sqrt{s}=400$ GeV and $\tht=30^{\circ}$. }
\end{center}
\end{figure}

\begin{figure}[ptb]
\begin{center}
\vskip -1.7cm
\input{dist3021.tex}
\vskip .6cm
\input{dist3022.tex}
\caption{ (a) The triple differential distribution including the  
QCD contribution 
for $\sqrt{s}=800$ GeV and $\tht=30^{\circ}$. 
(b) The fractional QCD contribution $\kappa(\tht,\thl,\phil)$ to the triple 
differential distribution 
for $\sqrt{s}=800$ GeV and $\tht=30^{\circ}$. }
\end{center}
\end{figure}

\begin{figure}[ptb]
\begin{center}
\vskip -1.7cm
\input{dist3031.tex}
\vskip .6cm
\input{dist3032.tex}
\caption{ (a) The triple differential distribution including the  
QCD contribution 
for $\sqrt{s}=1500$ GeV and $\tht=30^{\circ}$. 
(b) The fractional QCD contribution $\kappa(\tht,\thl,\phil)$ to the triple 
differential distribution 
for $\sqrt{s}=1500$ GeV and $\tht=30^{\circ}$. }
\end{center}
\end{figure}

We have not shown the plots for other values of $\tht$
because they do not illustrate anything new. We have also not discussed triple
distributions for the case of polarized beams for the sake of brevity.

\section{Conclusions}
To conclude, we have obtained in this paper analytic expressions for angular
distributions of leptons from top decay in $\ee \rightarrow \ttbar$, 
in the $\ee$
c.m. frame, including QCD corrections to order $\alpha_s$ in the soft-gluon 
approximation. The distributions are in a form which can be compared directly
with experiment. In particular, the single differential $\thl$ distribution
needs neither the reconstruction of the top momentum direction nor the top rest
frame. The triple differential distribution does need the top direction to be
reconstructed for the definition of the angles. However, the magnitude of the
top momentum need not be determined. In either case the
results do not depend on the choice of spin quantization axis. 

We find that the $\thl$ distributions are well described by rescaling the
zeroth order distributions by a factor $K$ which for $\sqrt{s}=400$ GeV 
is  roughly independent of
$\thl$, except for extreme values of $\thl$, for the case of right-handed
polarized electron beam. For other values of $\sqrt{s}$, it is a slowly varying
function of $\thl$. 
 
The triple distributions in $\tht$, $\thl$ and $\phil$ show an asymmetry
about $\phi_l = 180^{\circ}$, which is absent at tree level, and 
would be a good test of the QCD corrections.

It would be useful to carry out the hard-gluon corrections explicitly and check
if the soft-gluon approximation used here gives correct quantitative results.

\noindent 
{\bf Acknowledgements} I thank Werner Bernreuther, Arnd Brandenburg and V.
Ravindran for helpful discussions. I thank Ravindran also for clarifications
regarding the contribution of the imaginary parts of form factors. I also thank
Prof. P.M. Zerwas for hospitality in the DESY Theory Group, where this work was
begun.

\end{document}